# Synthesis and temperature-dependent photoluminescence of high density GeSe triangular nanoplate arrays on Si substrates

Xueyan Li,[a] Xi Zhang[*a], Xiaowei Lv,[b] Jun Pang,[a] Li Lei,[c] Yong Liu,[d] Yong Peng[b] and Gang Xiang[*a]

[a] *College of Physics, Sichuan University, Chengdu, 610064, China*

[b] *College of physical science and technology, Lanzhou University, Lanzhou 730000, China*

[c] *Institute of Atomic and Molecular Physics, Sichuan University, Chengdu 610065, China*

[d] *Analysis and Testing Center, Sichuan University, Chengdu, 610064, China*

\* *Corresponding authors: xizhang@scu.edu.cn and gxiang@scu.edu.cn*

**Abstract**

We have grown germanium selenide (GeSe) triangular nanoplate arrays (TNAs) with a high density ($3.82 \times 10^6$ / mm$^2$) on the Si (111) substrate using a simple thermal evaporation method. The thickness and trilateral lengths of a single triangular nanoplate were statistically estimated by atomic force microscopy (AFM) as 44 nm, 365 nm, 458 nm and 605 nm, respectively. Transmission electron microscopy (TEM) images and X-ray diffraction (XRD) patterns show that the TNAs were composed of single crystalline GeSe phase. The Se-related defects in the lattice were also revealed by TEM images and Raman vibration modes. Unlike previously reported GeSe compounds, the GeSe TNAs exhibited temperature-dependent photoluminescence (PL). In addition, not previously reported PL peak (1.25 eV) of the 44 nm thick TNAs at 5 K was in the gaps between those of GeSe monolayers (1.5 nm) and thin films (400 nm), revealing a close relationship between the PL peak and the thickness of GeSe. The high-density structure and temperature-dependent PL of the TNAs on the Si substrate may be useful for temperature controllable semiconductor nanodevices.

## 1. Introduction

Graphene is one of the most widely studied two-dimensional (2D) materials due to its fascinating mechanical, electronic and optical properties.[1-6] Because of the absence of bandgap in graphene, a great amount of effort has been put on finding alternative 2D semiconductors.[7-10] Recently, IV-VI composite semiconductors (such as GeS, GeSe, SnS and SnSe) have attracted extensive interests because of their natural layered crystalline structure and excellent semiconductor properties.[11-17] Additionally, due to the strong covalent bond within the layer and the weak van der Waals interaction between the layers, the suspended bond and surface state are eliminated, which makes the IV-VI materials possess chemical inert surface and considerable environmental stability.[18-21] Among IV-VI compound semiconductors, GeSe possesses a narrow band gap overlapping well with the solar spectrum, high carrier mobility, low toxicity and air stability, showing great potentials in fabrication of electronic and optoelectronic devices.[18, 21-25] Different GeSe structures including nanoparticles,[26] nanosheets (hexagonal, comb and irregular structures)[3, 12, 14, 18] and thin films[19, 25] were fabricated by various methods such as chemical vapor deposition and mechanical exfoliation.

There were a few reports investigating the photoluminescence (PL) properties of GeSe compounds. For example, Zhao *et al.* fabricated 1.5 nm thick GeSe monolayers using laser-thinning technology, and found that the PL spectra was temperature-independent in which the main peak was around 655 nm.[27-28] Chen *et al*. measured the PL spectra of 400 nm thick GeSe thin films (~ 267 layers) synthesized by thermal evaporation and found the main peak was around 1172 nm at room temperature.[24]

Obviously, there was a huge size gap between a monolayer thickness and a few hundred layers thickness and a very big difference between the PL spectra peaked at 655 nm and 1172 nm. Additionally, it was surprising that no temperature dependence of PL was found in GeSe yet, because PL is closely related to semiconductor band gap which is usually dependent on temperature. Therefore, it is important to fabricate GeSe structures with a thickness between a monolayer and a few hundred layers thickness and experimentally explore the thickness- and temperature-dependent PL properties.

In this report, we have fabricated high density and uniformly distributed GeSe triangular nanoplate arrays (TNAs) on Si (111) substrates by thermal evaporation. The thickness and trilateral lengths of a single triangular nanoplate were statistically estimated to be 44 nm, 365 nm, 458 nm and 605 nm, respectively. The microstructure and the PL of the TNAs were then studied. A weak temperature dependence of PL was observed in the GeSe TNAs and a novel PL main peak about 995 nm was found at 5 K. Discussions about band gap and activation energy of the GeSe TNAs were then presented.

## 2. Experimental

The samples were grown in a double-temperature zone corundum tube furnace system shown in Fig. 1(a). GeSe powders (99.9%) evenly spread (1 mm thick) on a piece of clean corundum plate (50 × 25 × 0.8 mm) located at the high-temperature zone. A piece of intrinsic Si (111) (10 × 10 × 0.5 mm) was placed in the low-temperature zone as the substrate. The corundum tube was first evacuated by a mechanical rotary pump before evaporation and then a carrier gas (5% $H_2$ + 95% Ar) was flowed through at 30 standard cubic centimeters per minute (sccm) at 5 Torr during the whole growth process. Then GeSe powders were heated at 470 °C to form GeSe vapors.[19] The vapors were transported over the Si substrate along with the carrier gas flowing through the chamber. After 1 h thermal evaporation, GeSe products were formed on the Si substrate. In order to optimize the growth conditions, different substrate temperatures (50 °C, 100 °C and 150 °C) were tried. After the growth process, the furnace was naturally cooled down to room temperature. The schematic diagram of the growth process was shown in Fig. 1(b).

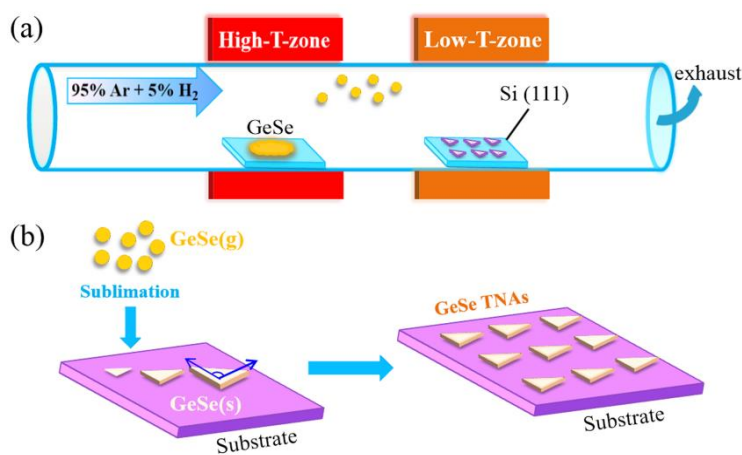

**Fig. 1** (a) Schematic of experimental setup. (b) The growth process of GeSe TNAs on Si (111) substrate.

The samples' morphologies were studied with atomic force microscopy (AFM, Benyuan, CSPM5500) and transmission electron microscopy (TEM, FEI, Tecnai-F30). The structural and chemical characterizations were conducted by TEM equipped with an energy dispersive X-ray spectroscopy (EDX, AMETEK). The crystalline nature was characterized by high-resolution TEM (HRTEM) and corresponding Fourier transformation (FFT). X-ray diffraction (XRD) patterns were recorded on a diffractometer (Fangyuan, DX-2500) equipped with a Cu Kα radiation (λ = 1.54056 Å). X-ray photoelectron spectroscopy (XPS, Thermo Fisher, K-Alpha) with Al Kα radiation of energy 1486.8 eV was used to analyze the composition and chemical state. Raman spectra was collected from liquid nitrogen temperature to room temperature using a solid-state laser with a wavelength of 532 nm as the excitation source. The PL properties were studied by a fluorescence spectrometer (Princeton Instruments, SP2500i) in the temperature range of 5 K to 240 K using a He-Cd laser in the spectral range of 813-1300 nm excited at a wavelength of 405 nm.

**3. Results and discussion**

Fig. 2(a-c) shows 2D AFM images of the synthesized GeSe products on the Si (111) substrate heated at different temperatures (50 °C, 100 °C and 150 °C). At 50 °C, the products were composed of small irregular dots sparsely distributed on the substrate. At 100 °C, a large number of densely and uniformly distributed triangular nanoplate arrays were formed. At 150 °C, the triangular nanoplates became blurred and non-uniformly distributed. Obviously, the substrate temperature played a key role in the formation of the GeSe TNAs, and it turns out that 100 °C heating provided appropriate thermal energies for the GeSe molecules to form high quality TNAs. We also tried different GeSe source temperatures (350 °C and 530 °C) other than 470 °C, and found that either no product (350 °C) or only Ge particles (530 °C) synthesized on the Si substrate. So we will focus on the results of the TNAs grown at the soure temperature of 470 °C and the substrate temperature of 100 °C.

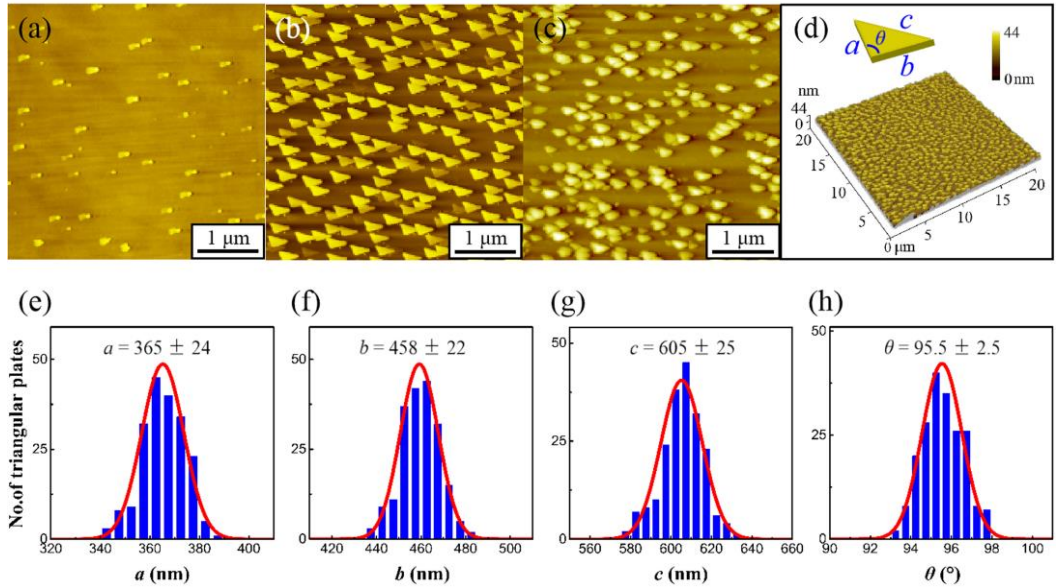

**Fig. 2** AFM 2D images of GeSe grown on Si (111) substrate at different temperatures (a) 50 °C, (b) 100 °C and (c) 150 °C. (d) AFM 3D image corresponding to (b), inset is the diagram of a triangular plate model. Length distribution of (e) *a*, (f) *b* and (g) *c*. (h) Angle *θ* between *a* and *b* of the triangular nanoplates.

Fig. 2(d) was three-dimensional (3D) AFM image of GeSe TNAs corresponding to Fig. 2(b), it shows that the thickness of a single triangular nanoplate was ~ 44 nm. The average density of triangular nanoplates on the substrate was measured by AFM as ~ $3.82 \times 10^6$ / mm$^2$. In Fig. 2(d), an ideal triangular plate model was established to characterize the basic structural parameters of the GeSe triangular nanoplate, where the edge lengths were represented by *a*, *b* and *c*, respectively, and the angle between *a* and *b* was represented by *θ*. Fig. 2(e-h) show the statistical distribution curves of the measured parameters *a*, *b*, *c* and *θ* of 200 randomly selected triangular nanoplates. By fitting the experimental curves with Gaussian distribution function, the most probable values of *a*, *b*, *c* and *θ* were obtained as 365 ± 24 nm, 458 ± 22 nm, 605 ± 25 nm and

95.5 ± 2.5 °, respectively.

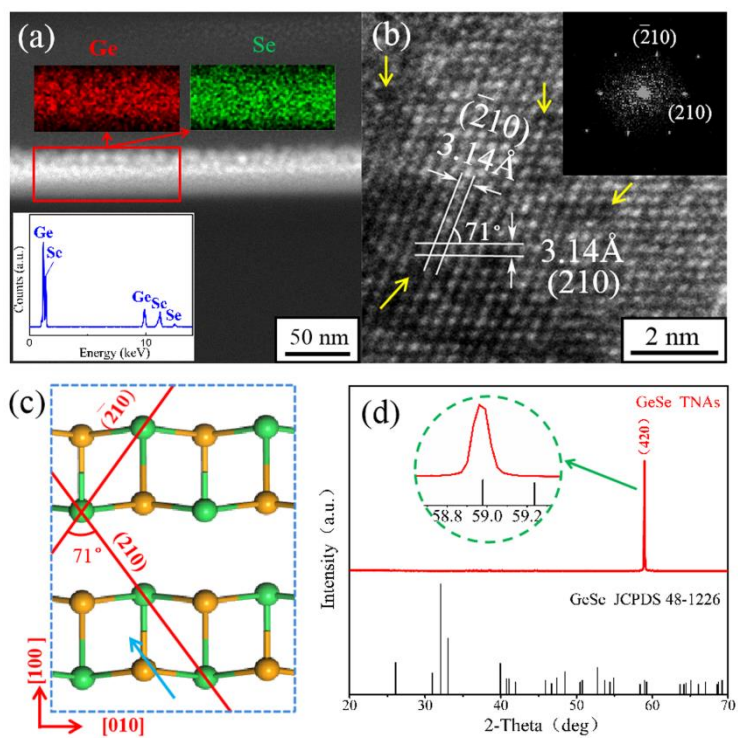

**Fig. 3** (a) Annular dark-field STEM image of the cross-section of the TNAs and corresponding EDS mapping images of Ge and Se elements. Inset is a representative EDS spectrum. (b) The HRTEM image of the area marked by red rectangular frame in (a) as well as the corresponding FFT image. The yellow arrows indicate the locations of the defects. (c) Crystallographic view of GeSe lattice on the (001) plane depicting the crystal facets of (210) and ($\bar{2}$10) matching the lattice image in (b). Green balls and yellow balls represent Ge and Se atoms, respectively. (d) XRD pattern of the GeSe TNAs on Si substrate.

The structure and chemical composition of the TNAs were investigated by TEM. An annular dark-field (ADF) STEM image of the cross-section of the TNAs was shown in Fig. 3(a). A representative EDS spectrum and a region scanning profile of Ge and Se elements indicate that the main ingredients of the TNAs were Ge and Se and the atomic ratio of Ge: Se was approximately 1:1. The region scanning profile confirmed the Ge and Se uniformly distributed in the triangular nanoplate. The HRTEM image in Fig. 3(b) reveals that the GeSe triangular nanoplate was a single crystalline structure, and the exposed crystalline facets in the lattice were (210) and ($\bar{2}$10) with corresponding crystalline plane spacing of 3.14 Å. The defective sites marked by the yellow arrows in the HRTEM image and the blurry halo shown in the corresponding FFT image indicate there is a small amount of defective phases presented in the nanoplates. Further Raman studies also reveal existence of the defects in the TNAs, which will be discussed later. Fig. 3(c) shows a schematic diagram of 2D GeSe crystalline structure, which depicts the crystalline facets of (210) and ($\bar{2}$10) in Fig. 3 (b). The XRD patterns of the TNAs in Fig. 3(d) show that the diffraction peak corresponding to the crystalline facet (420) could be indexed to the orthorhombic GeSe with cell unit of $a$ = 10.84 Å, $b$ = 3.834 Å, $c$ = 4.39 Å (JCPDS No. 48−1226, Pnma), consistent with the HRTEM result.

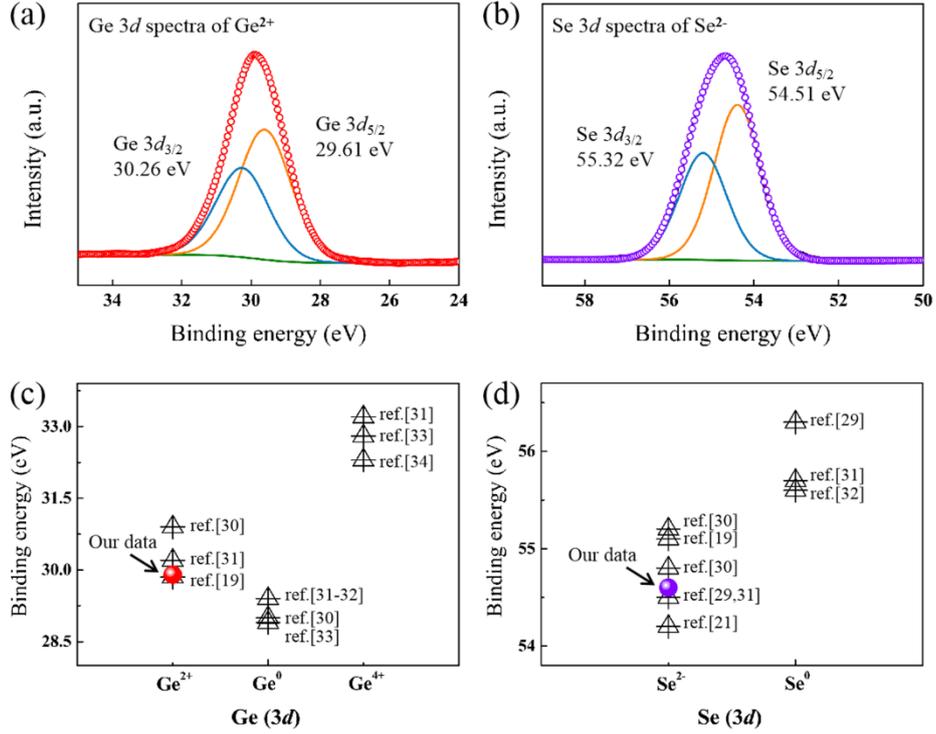

**Fig. 4** XPS spectra of (a) Ge 3*d* and (b) Se 3*d* in the GeSe TNAs. The binding energies of (c) $Ge^{2+}$ and (d) $Se^{2-}$ detected by XPS in our work (circles) and in the literature (triangles).

The bonding states in the GeSe TNAs were analyzed by XPS. The spectra of the TNAs exhibit the peaks of Ge 3*d* around 29.9 eV and Se 3*d* around 54.6 eV, as shown in Fig. 4(a) and (b). The binding energies of Ge 3*d* and Se 3*d* in our measurements were in good agreement with those of GeSe compounds in literatures,[19, 29] as shown in Fig. 4(c) and (d), indicating that the most probable valence states of Ge and Se were +2 and -2 in the TNAs, and $Se^0$, $Ge^0$ and $Ge^{4+}$ were not detected within the limitation of XPS resolution.

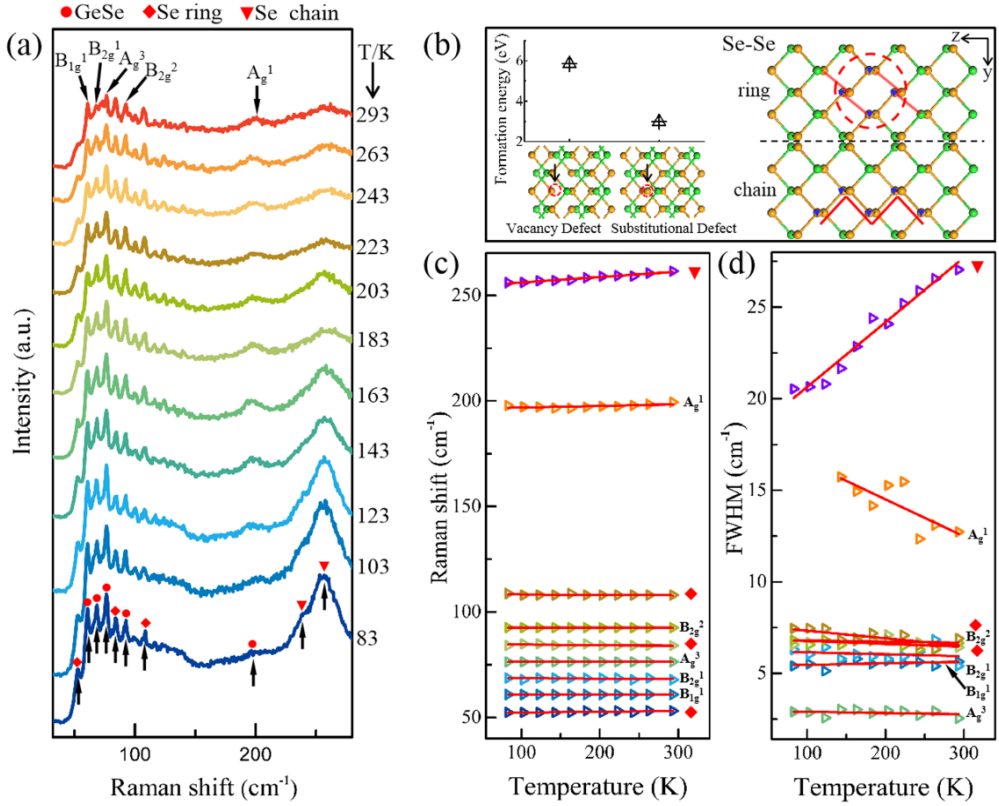

**Fig. 5** (a) Raman spectra collected from the GeSe TNAs under 532 nm laser light excitation at the temperatures between 83 K and 293 K. (b) Left: the formation energies of vacancy defect and substitutional defect on the Ge site in GeSe by *ab initio* calculations, respectively. Right: the structures of $Se_8$ molecule (ring) and polymer-like structure (chain). The variation of (c) peak position and (d) full width at half maximum (FWHM) as the function of temperature.

Fig. 5(a) shows the typical Raman spectra collected from the GeSe TNAs under 532 nm laser excitation at the temperatures between 83 K and 293 K. The Raman peak signal was weak. The similar phenomenon has been observed in laser thinned GeSe nanosheets, which was attributed to the small thickness and the defects of the sample.[27,35] Fig. 5(a) exhibited ten discernible Raman peaks located at 52 $cm^{-1}$, 61 $cm^{-1}$, 69 $cm^{-1}$,

76 cm$^{-1}$, 84 cm$^{-1}$, 93 cm$^{-1}$, 109 cm$^{-1}$, 197 cm$^{-1}$, 239 cm$^{-1}$ and 256 cm$^{-1}$, respectively. According to previous Raman studies on GeSe compounds,[27, 35-36] these peaks correspond to phonon vibrations of Ge-Se bonds and Se-related defects. The ten peaks could be well divided into two groups. One group includes five peaks located at 61 cm$^{-1}$, 69 cm$^{-1}$, 76 cm$^{-1}$, 93 cm$^{-1}$ and 197 cm$^{-1}$, which could be identified as $B_{1g}^1$, $B_{2g}^1$, $A_g^3$, $B_{2g}^2$ and $A_g^1$ vibration modes of the Ge-Se bonds,[35-37] respectively. The other group includes the other five peaks at 52 cm$^{-1}$, 84 cm$^{-1}$, 109 cm$^{-1}$, 239 cm$^{-1}$ and 256 cm$^{-1}$, which were assigned to Se-related defects: the first three peaks correspond to the vibration modes of selenium octacyclic (Se$_8$) molecule,[38] and the last two peaks correspond to the polymer-like structure.[39] The deviation between the experimental and theoretical values of each Raman peak position was not more than 5 cm$^{-1}$.

The Se$_8$ molecule and polymer-like defects in our samples were probably resulted from the absence or disordered arrangement of lattice atoms accidentally occurred during the growth process. One example of the possible formations of these two Se-related structures was shown in Fig. 5(b), in which Se$_8$ molecule (ring) and polymer-like structure (chain) were formed by Se and Se interatomic bonding after Ge atoms replaced by Se ones. We did not use vacancy defects to construct the Se-related structures in Fig. 5(b) because our *ab initio* calculations showed that the formation energy of substitutional defect on the Ge site in GeSe lattice was less than that of vacancy defect, suggesting the substitutional structure was energetically more preferable in the nanoplates. The Raman peak signal at 256 cm$^{-1}$ was relatively strong in our samples, indicating that the polymer-like structure dominates in the defects of the GeSe TNAs.

The Raman peaks were fitted by Gaussian distribution functions and the temperature-dependent peak position and full width at half maximum (FWHM) were plotted in Fig. 5(c-d). The Raman peak position and FWHM of each vibration modes vary almost linearly with temperature and no sudden change occurs, which indicates that no structural phase transitions happen below 293 K. However, the slopes of the curves are different, indicating that the influence of temperature on different vibration modes was different.

The PL properties of the GeSe TNAs were studied under 405 nm laser excitation at different temperatures. Fig. 6(a) illustrates that the samples were capable of emitting light under the excitation below 220 K. Unlike previous reports by other groups,[24, 27-28] the temperature-dependent PL was observed in our GeSe samples: the luminescence peak position was around 995 nm (1.25 eV) at 5 K and increased to 1017 nm (1.22 eV) at 220 K. In addition, the PL peak position of the TNAs was also significantly different from previously reported results. For comparison, we plotted the experimental PL peak positions of 1.5 nm thick GeSe monolayers,[27-28] 400 nm thick thin films[24] and our 44 nm thick TNAs in Fig. 6(d), which shows that the TNAs exhibit a luminescence peak in the gap between those of the GeSe monolayers and thin films. This reveals that the main PL peak is dependent on the thickness of GeSe compounds, which is consistent with the layer-dependent behavior of the band gap and PL by *ab initio* calculations.[40-41]

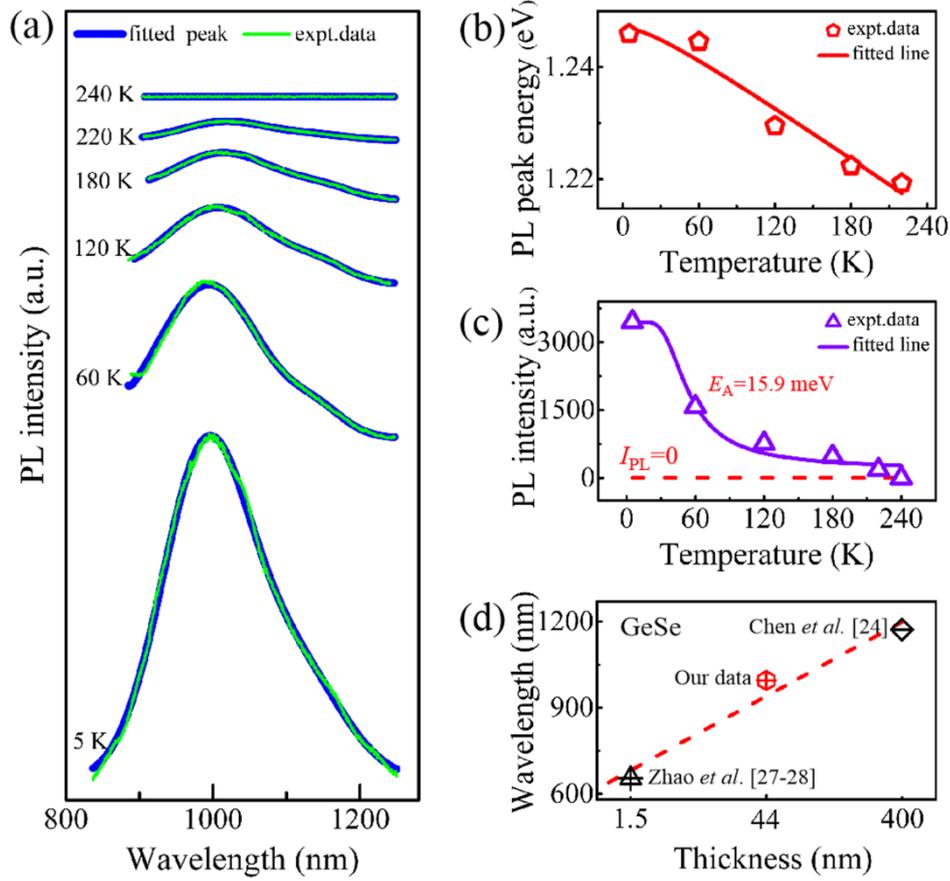

**Fig. 6** (a) The temperature-dependent PL spectra of the GeSe TNAs. (b) The PL peak energy as a function of temperature. (c) The PL peak intensity as a function of temperature. The value of $E_A$ = 15.9 meV was obtained in the GeSe TNAs by fitting the experimental data. (d) The experimental value of the main PL peaks of GeSe monolayers, thin films and our triangular nanoplates.

The PL spectra were fitted with a Gaussian distribution function, and the temperature-dependent PL peak energy and intensity were plotted in Fig. 6(b) and (c). The PL peak energy as a function of temperature can be well described by Varshni equation[42]

$$E(T) = E(0) - \frac{\alpha T^2}{(\beta + T)} \quad (1),$$

where $E(0)$ represents the band gap at $T = 0$ K, $\alpha$ is the temperature coefficient, and $\beta$ is a constant related to the material. The fitting to Eq. (1) shows that $E(0) = 1.247$ eV, $\alpha = 0.16$ meVK$^{-1}$ and $\beta = 34.89$ K. The PL intensity decreased with increasing temperature, since more carriers were thermally excited to higher energy states at higher temperature and then diffused away to non-radiative recombination centers related to the defects and recombined there.[43-44] The curve in Fig. 6(c) shows a complete exponential decay process, which can be well expressed by Arrhenius equation[45]

$$I(T) = \frac{I_0}{1+C\exp(-\frac{E_A}{k_BT})} \quad (2),$$

where $I_0$ is a normalizing factor, $C$ is a constant related to radiative and non-radiative recombination, $E_A$ is the activation energy considering the non-radiative recombination process, and $k_B$ is the Boltzmann constant. The value of $E_A$ can be obtained by fitting the experimental data of intensity ($I$) as a function of temperature ($T$) according to Eq. (2). As shown in Fig. 6(c), the $E_A$ value of 15.9 meV was obtained in the GeSe triangular nanoplates. This value is the same order of magnitude as those in other II-VI and IV-VI compounds, such as CdS nanobelts[46] and PbS quantum dots[47]. The temperature-dependent novel PL wavelength and low PL activation energy of the GeSe TNAs may be useful for semiconductor nanodevices.

## 4. Conclusion

In conclusion, we have grown GeSe triangular nanoplate arrays with a high density of $3.82 \times 10^6$ / mm$^2$ on the Si substrate. The structures and temperature-dependent PL properties of the TNAs were investigated. The GeSe TNAs can emit not previously reported intense infrared light under photon excitation, whose wavelength was around 995 nm at 5 K and activation energy was 15.9 meV. The temperature-dependent PL characteristics of the GeSe TNAs on the Si substrate may be utilized to design GeSe-based semiconductor devices, such as the temperature-controllable infrared light generators or photo-thermal sensors.

**Conflicts of interest**

There are no conflicts to declare.

**Acknowledgements**

This work was supported by National Key R&D Program of China through Grant No. 2017YFB0405702 and by the Natural Science Foundation of China (NSFC) through Grant No. 51672179.

**References**

1 Tan, D.; Lim, H. E.; Wang, F.; Mohamed, N. B.; Mouri, S.; Zhang, W.; Miyauchi, Y.; Ohfuchi, M.; Matsuda, K., Anisotropic optical and electronic properties of two-dimensional layered germanium sulfide. *Nano Research* **2016,** *10* (2), 546-555.

2 Ramasamy, P.; Kwak, D.; Lim, D.-H.; Ra, H.-S.; Lee, J.-S., Solution synthesis of GeS and GeSe nanosheets for high-sensitivity photodetectors. *Journal of Materials Chemistry C* **2016,** *4* (3), 479-485.

3 Liu, J.; Zhou, Y.; Lin, Y.; Li, M.; Cai, H.; Liang, Y.; Liu, M.; Huang, Z.; Lai, F.; Huang, F.; Zheng, W., Anisotropic Photoresponse of the Ultrathin GeSe Nanoplates Grown by Rapid Physical Vapor Deposition. *ACS Appl Mater Interfaces* **2019,** *11* (4), 4123-4130.

4 Levendorf, M. P.; Kim, C. J.; Brown, L.; Huang, P. Y.; Havener, R. W.; Muller, D. A.; Park, J., Graphene and boron nitride lateral heterostructures for atomically thin circuitry. *Nature* **2012,** *488* (7413), 627-32.

5 Novoselov, K. S.; Geim, A. K.; Morozov, S. V.; Jiang, D.; Zhang, Y.; Dubonos, S. V.; Grigorieva, I. V.; Firsov, A. A., Electric field effect in atomically thin carbon films. *science* **2004,** *306* (5696), 666-669 %@ 0036-8075.

6 Fan, Z.-Q.; Jiang, X.-W.; Wei, Z.; Luo, J.-W.; Li, S.-S., Tunable Electronic Structures of GeSe Nanosheets and Nanoribbons. *The Journal of Physical Chemistry C* **2017,** *121* (26), 14373-14379.

7 Mak, K. F.; Lee, C.; Hone, J.; Shan, J.; Heinz, T. F., Atomically thin MoS(2): a new direct-gap semiconductor. *Phys Rev Lett* **2010,** *105* (13), 136805.

8 Wang, Q. H.; Kalantar-Zadeh, K.; Kis, A.; Coleman, J. N.; Strano, M. S., Electronics and optoelectronics of two-dimensional transition metal dichalcogenides. *Nat Nanotechnol* **2012,** *7* (11), 699-712.

9 Li, L.; Yu, Y.; Ye, G. J.; Ge, Q.; Ou, X.; Wu, H.; Feng, D.; Chen, X. H.; Zhang, Y., Black phosphorus field-effect transistors. *Nat Nanotechnol* **2014,** *9* (5), 372-7.

10 Novoselov, K. S.; Fal'ko, V. I.; Colombo, L.; Gellert, P. R.; Schwab, M. G.; Kim, K., A roadmap for graphene. *Nature* **2012,** *490* (7419), 192-200.

11 Li, L.; Chen, Z.; Hu, Y.; Wang, X.; Zhang, T.; Chen, W.; Wang, Q., Single-layer single-crystalline SnSe nanosheets. *J Am Chem Soc* **2013,** *135* (4), 1213-6.

12 Mukherjee, B.; Cai, Y.; Tan, H. R.; Feng, Y. P.; Tok, E. S.; Sow, C. H., NIR Schottky photodetectors based on individual single-crystalline GeSe nanosheet. *ACS Appl Mater Interfaces* **2013,** *5* (19), 9594-604.

13 Vaughn, D. D.; Patel, R. J.; Hickner, M. A.; Schaak, R. E., Single-crystal colloidal nanosheets of GeS and GeSe. *Journal of the American Chemical Society* **2010,** *132* (43), 15170-15172 %@ 0002-7863.

14 Yoon, S. M.; Song, H. J.; Choi, H. C., p-type semiconducting GeSe combs by a vaporization-condensation-recrystallization (VCR) process. *Adv Mater* **2010,** *22* (19), 2164-7.

15 Lu, J.; Guo, L.; Xiang, G.; Nie, Y.; Zhang, X., Electronic and Magnetic Tunability of SnSe Monolayer via Doping of Transition-Metal Atoms. *Journal of Electronic Materials* **2019**.

16 Ma, Q.; Zhang, X.; Yang, D.; Xiang, G., Impact of side passivation on the electronic structures and optical properties of GeSe nanobelts. *Superlattices and Microstructures* **2019,** *125*, 365-370.

17 Cai, X.; Luo, J.; Zhang, X.; Xiang, G., The electronic structures and optical properties of light-element atom adsorbed SnSe monolayers. *Materials Research Express* **2018,** *5* (3), 035013.

18 Zhou, X.; Hu, X.; Jin, B.; Yu, J.; Liu, K.; Li, H.; Zhai, T., Highly Anisotropic GeSe Nanosheets for Phototransistors with Ultrahigh Photoresponsivity. *Adv Sci (Weinh)* **2018,** *5* (8), 1800478.

19 Xue, D. J.; Liu, S. C.; Dai, C. M.; Chen, S.; He, C.; Zhao, L.; Hu, J. S.; Wan, L. J., GeSe Thin-Film Solar Cells Fabricated by Self-Regulated Rapid Thermal Sublimation. *J Am Chem Soc* **2017,** *139* (2), 958-965.

20 von Rohr, F. O.; Ji, H.; Cevallos, F. A.; Gao, T.; Ong, N. P.; Cava, R. J., High-Pressure Synthesis and Characterization of beta-GeSe-A Six-Membered-Ring Semiconductor in an Uncommon Boat Conformation. *J Am Chem Soc* **2017,** *139* (7), 2771-2777.

21 Xue, D. J.; Tan, J.; Hu, J. S.; Hu, W.; Guo, Y. G.; Wan, L. J., Anisotropic photoresponse properties of single micrometer-sized GeSe nanosheet. *Adv Mater* **2012,** *24* (33), 4528-33.

22 Ma, D.; Zhao, J.; Wang, R.; Xing, C.; Li, Z.; Huang, W.; Jiang, X.; Guo, Z.; Luo, Z.; Li, Y.; Li, J.; Luo, S.; Zhang, Y.; Zhang, H., Ultrathin GeSe Nanosheets: From Systematic Synthesis to Studies of Carrier Dynamics and Applications for a High-Performance UV-Vis Photodetector. *ACS Appl Mater Interfaces* **2019,** *11* (4), 4278-4287.

23 Vaughn, D.; Sun, D.; Levin, S. M.; Biacchi, A. J.; Mayer, T. S.; Schaak, R. E., Colloidal Synthesis and Electrical Properties of GeSe Nanobelts. *Chemistry of Materials* **2012,** *24* (18), 3643-3649.

24 Chen, B.; Ruan, Y.; Li, J.; Wang, W.; Liu, X.; Cai, H.; Yao, L.; Zhang, J. M.; Chen, S.; Chen, G., Highly oriented GeSe thin film: self-assembly growth via the sandwiching post-annealing treatment and its solar cell performance. *Nanoscale* **2019,** *11* (9), 3968-3978.

25 Dhanabalan, S. C.; Ponraj, J. S.; Zhang, H.; Bao, Q., Present perspectives of broadband photodetectors based on nanobelts, nanoribbons, nanosheets and the emerging 2D materials. *Nanoscale* **2016,** *8* (12), 6410-34.

26 Wang, K.; Huang, D.; Yu, L.; Feng, K.; Li, L.; Harada, T.; Ikeda, S.; Jiang, F., Promising GeSe Nanosheet-Based Thin-Film Photocathode for Efficient and Stable Overall Solar Water Splitting. *ACS Catalysis* **2019,** *9* (4), 3090-3097.

27 Zhao, H.; Mao, Y.; Mao, X.; Shi, X.; Xu, C.; Wang, C.; Zhang, S.; Zhou, D., Band Structure and Photoelectric Characterization of GeSe Monolayers. *Advanced Functional Materials* **2018,** *28* (6), 1704855.

28 Mao, Y.; Mao, X.; Zhao, H.; Zhang, N.; Shi, X.; Yuan, J., Enhancement of photoluminescence efficiency in GeSe ultrathin slab by thermal treatment and annealing: experiment and first-principles molecular dynamics simulations. *Sci Rep* **2018,** *8* (1), 17671.

29 Naveau, A.; Monteil-Rivera, F.; Guillon, E.; Dumonceau, J., Interactions of Aqueous Selenium (−II) and (IV) with Metallic Sulfide Surfaces. *Environmental Science & Technology* **2007,** *41* (15), 5376-5382.

30 Shalvoy, R. B.; Fisher, G. B.; Stiles, P. J., Bond ionicity and structural stability of some average-valence-five materials studied by x-ray photoemission. *Physical Review B* **1977,** *15* (4), 1680-1697.

31 Crist, B. V. "BE lookup table for signals from elements and common chemical species." Handbook of Monochromatic XPS Spectra: The Elements of Native Oxides 1 (1999): 77-358.

32 Chastain, Jill. "Handbook of X-ray photoelectron spectroscopy." Perkin-Elmer Corporation 40 (1992): 221.

33 Mukherjee, S.; Das, K.; Das, S.; Ray, S. K., Highly Responsive, Polarization Sensitive, Self-Biased Single $GeO_2$-Ge Nanowire Device for Broadband and Low Power Photodetectors. *ACS Photonics* **2018,** *5* (10), 4170-4178.

34 Zhou, X.; Hu, X.; Zhou, S.; Zhang, Q.; Li, H.; Zhai, T., Ultrathin 2D $GeSe_2$ Rhombic Flakes with High Anisotropy Realized by Van der Waals Epitaxy. *Advanced Functional Materials* **2017,** *27* (47), 1703858.

35 Ye, Y.; Guo, Q.; Liu, X.; Liu, C.; Wang, J.; Liu, Y.; Qiu, J., Two-Dimensional GeSe as an Isostructural and Isoelectronic Analogue of Phosphorene: Sonication-Assisted Synthesis, Chemical Stability, and Optical Properties. *Chemistry of Materials* **2017,** *29* (19), 8361-8368.

36 Taube, A.; Łapińska, A.; Judek, J.; Wochtman, N.; Zdrojek, M., Temperature induced phonon behaviour in germanium selenide thin films probed by Raman spectroscopy. *Journal of Physics D: Applied Physics* **2016,** *49* (31), 315301.

37 Gashimzade, F. M.; Guseinova, D. A.; Jahangirli, Z. A.; Nizametdinova, M. A., Ab initio calculation of vibrational spectra of orthorhombic IV–VI layered crystals. *Physics of the Solid State* **2013,** *55* (9), 1802-1807.

38 Gorman, M., and S. A. Solin. "Transmission Raman and depolarization spectra of bulk a-Se from 13 to 300 cm− 1." Solid State Communications 18.11-12 (1976): 1401-1404.

39 Poborchii, V. V.; Kolobov, A. V.; Tanaka, K., An in situ Raman study of polarization-dependent photocrystallization in amorphous selenium films. *Applied Physics Letters* **1998,** *72* (10), 1167-1169.

40 Shi, G.; Kioupakis, E., Anisotropic Spin Transport and Strong Visible-Light Absorbance in Few-Layer SnSe and GeSe. *Nano Lett* **2015,** *15* (10), 6926-31.

41 Song, X.; Zhou, W.; Liu, X.; Gu, Y.; Zhang, S., Layer-controlled band alignment, work function and optical properties of few-layer GeSe. *Physica B: Condensed Matter* **2017,** *519*, 90-94.

42 Varshni, Y P., Temperature dependence of the energy gap in semiconductors. physica, 1967, 34(1): 149-154.

43 Nan, H.; Wang, Z.; Wang, W.; Liang, Z.; Lu, Y.; Chen, Q.; He, D.; Tan, P.; Miao, F.; Wang, X., Strong photoluminescence enhancement of MoS2 through defect engineering and oxygen bonding. *ACS nano* **2014,** *8* (6), 5738-5745 % @ 1936-0851.

44 Platt, A. D.; Kendrick, M. J.; Loth, M.; Anthony, J. E.; Ostroverkhova, O., Temperature dependence of exciton and charge carrier dynamics in organic thin films. *Physical Review B* **2011,** *84* (23).

45 Leroux, M.; Grandjean, N.; Beaumont, B.; Nataf, G.; Semond, F.; Massies, J.; Gibart, P., Temperature quenching of photoluminescence intensities in undoped and doped GaN. Journal of Applied Physics, 1999, 86(7): 3721-3728.

46 Liu, B.; Chen, R.; Xu, X. L.; Li, D. H.; Zhao, Y. Y.; Shen, Z. X.; Xiong, Q. H.; Sun, H. D., Exciton-Related Photoluminescence and Lasing in CdS Nanobelts. *The Journal of Physical Chemistry C* **2011,** *115* (26), 12826-12830.

47 Wang, J.; Mandelis, A.; Sun, Q.; Li, B.; Gao, C., Temperature- and Size-Dependent Exciton Dynamics in PbS Colloidal Quantum Dot Thin Films Using Combined Photoluminescence Spectroscopy and Photocarrier Radiometry. *The Journal of Physical Chemistry C* **2018,** *122* (10), 5759-5766.